\documentclass{PNASTWO}

\usepackage{graphicx}
\usepackage{rotate}
\usepackage{epsfig}
\usepackage{cite}

\usepackage{amssymb,amsfonts,amsmath}
\providecommand{\e}[1]{\ensuremath{\times 10^{#1}}}

\begin{document}



\conflictofinterest{Conflict of interest footnote placeholder}

\track{This paper was submitted directly to the
PNAS office.}


\footcomment{Abbreviations: email, email network of Universitat Rovira i Virgili, Tarra\-gona, Spain;
PGP, 'Pretty-Good-Privacy' encryption network; IH, intracommunity heterogeneity;
IC, intercommunity connectivity}


\title{Mesoscopic structure conditions the emergence of cooperation on social networks}

\author{Sergi Lozano\affil{1}{Departament d'Enginyeria Inform\`atica i
Matem\`atiques,
Universitat Rovira i Virgili,
Av.\ Pa\"\i sos Catalans, 26, 43007 Tarragona, Spain}, Alex Arenas\affil{1}{}\affil{2}{Instituto de Biocomputaci\'on y F\'\i sica de Sistemas Complejos (BIFI),
Universidad de Zaragoza, 50009 Zaragoza, Spain}%
\and Angel S\'anchez \thanks{<To whom correspondence should be addressed. E-mail:
anxo@math.uc3m.es}
\affil{3}{Grupo Interdisciplinar de Sistemas Complejos (GISC),
Departamento de Matem\'aticas, Universidad Carlos III de Madrid, 28911 Legan\'es, Spain}\affil{2}{}
}

\contributor{Submitted to Proceedings of the National Academy of Sciences
of the United States of America}

\maketitle

\begin{article}

\begin{abstract}
We study the evolutionary Prisoner's Dilemma on two social networks
obtained from actual relational data.
We find very different cooperation levels on each of them that can
not be easily understood in terms of global statistical properties
of both networks. We claim that the result can be understood
at the mesoscopic scale, by studying the community structure of
the networks. We explain the
dependence of the cooperation level on the temptation parameter in
terms of the internal structure of the communities and their
interconnections. We then test our results on community-structured,
specifically designed artificial networks, finding perfect agreement
with the observations in the real networks.
Our results support the conclusion that studies of evolutionary
games on model networks and their interpretation in terms of global
properties  may not be sufficient to study specific, real social systems.
In addition, the community perspective
may be helpful to interpret
the origin and behavior of existing networks as well as to design
structures that show resilient cooperative behavior.
\end{abstract}




\section{Introduction}
\dropcap{T}he emergence and survival of cooperation in adverse environments
has been, for a long time, a challenging problem for scholars in
disciplines as diverse as biology, sociology or economics
\cite{Dugatkin, Hammerstein, Gintis1}. While some partial
answers have been advanced
in the last forty years \cite{Nowak1}, cooperation among unrelated
individuals is far from understood. Social dilemmas, situations in
which individual rationality leads to situations in which everyone
is worse off, are a prominent example of this conundrum
\cite{Kollock}. Within the general framework of evolutionary game
theory, which is particularly well suited to study this problem, the
Prisoner's Dilemma (PD) is a paradigmatic setting to
capture the paradox of altruism persistence against
short-term benefits of egoism. In this game two players choose
between cooperation (C) or defection (D), the payoffs for the two actions
being as in the following payoff matrix (payoffs for the row player are
indicated):
\begin{equation}
\label{matrixPD}
\begin{array}{ccc}
 & \mbox{ }\, C & \!\!\!\!\!\! D \\
\begin{array}{c} \!\!\! C \\ \!\!\! D \end{array} & \left(\begin{array}{c} 1 \\ b
\end{array}\right. & \left.\begin{array}{c} 0 \\ \epsilon
\end{array}\right). \end{array} \end{equation}
Relations between different possible payoffs follow the rule $b > 1
> \epsilon > 0$, that immediately poses the dilemma:
While the rational choice is to defect, it leads to a highly inefficient outcome as
compared to that obtained by two cooperators.

Among the plethora of studies devoted to this issue, a particularly
important and fruitful one is the modeling of the population as a
set of non-rational, learning agents that interact locally
\cite{Nowak2,Huberman,Nowak3,Eshel,Kirchkamp,Hauert} (see \cite{Szabo}
for a very recent review). Locality
is introduced in the model through a network on which agents
are placed. These agents then play the game only with their
neighbors (in neighborhoods that can be defined in different ways)
instead of interacting with all other agents. Learning is introduced
through imitation: after a round of games has been carried through
the whole lattice, agents look at their neighbors and choose the
strategy that has led to the highest payoff before proceeding to the
next round of games. With these two ingredients, namely locality and
imitation, it is generally
observed\cite{Nowak2,Hauert,nosotros} that states in which a sizeable part of the population
cooperates emerge (at least for values of $b$ not too close to $2$),
the mechanism for this emergence being the
formation of clusters of cooperators that can successfully
outcompete defectors.

Naturally,
the question arises as to whether this mechanism for the emergence
of cooperation appears 
also in real social networks \cite{Wasserman}. As a first step to answer this
question, some authors have focused their interest on the influence
of certain structural features that have been observed in real
networks on the evolution of cooperation, such as the small-world
phenomenon \cite{Abramson} or the scale-free character of the degree
distribution \cite{Santos}. A general conclusion of this research is that the
inhomogeneity of the degree distribution plays a central role on
this issue, and that it may favor the emergence of cooperation.
However, none of these studies deals with true social networks, as
they are all based on different types of artificial models.
To our knowledge, there is only one paper about the PD on real
social networks \cite{Holme}, but its point
of view is dynamical and unrelated to the present one.
Therefore, our research is a first
attempt to understand the relevance of considering empirical social
networks as a support for the local interactions in the framework of
imitation models.

\section{Experimental measurements on real networks}

\subsection{Datasets}
For our research we have used two
social substrates obtained by sampling real relational data.
We have chosen these substrates instead of other social network data available, such as the IMDB network for actor collaboration in movies or the scientific collaborating
networks, because their links are defined through true personal exchanges. In contrast, these other public data are bipartite networks, where links are defined by joining the collaboration framework (movies, research projects, articles, etc.)  which does not necessarily imply mutual interactions.
Our first substrate is a social network obtained from the email
traffic between members of University Rovira i Virgili (in
Tarragona, Spain; email network from now on),
where nodes represent individual email addresses
and undirected links between two nodes indicate bidirectional
communication (at least one email in each direction)
\cite{Guimera}. Our second real social substrate
consists of nodes representing users of the "Pretty-Good-Privacy" encryption
algorithm (PGP network, from now on),
while links trace trust relationships between those persons
who sign each other's public keys \cite{Boguña}. For a comparison of some
of their statistical properties see Table \ref{tab:nwk}.

\subsection{Dynamics}
Our simulations of the PD over all the networks (both email and PGP,
as well as on the models to be introduced below) follow strictly the
rules in \cite{Nowak2,Nowak3}, namely:
\begin{itemize}
    \item Initial strategies of agents are assigned randomly with the same probability to be C or D (we have
checked that other choices for the initial fraction of C or D lead
to similar results).
    \item The game is played between each pair of neighbors, and payoffs
    are accrued according to Eq.\ (\ref{matrixPD}), with $\epsilon=0$ although we checked that its value (being small, e.g. $0.01$) does not affect the results.
    \item Accumulated payoffs of all agents are computed by adding up the results of the games with their neighbors in the present turn.
    \item In the next round, every agent imitates the strategy of the most successful agent in her neighborhood (randomly selected if
there are two or more agents with the same payoff), after which payoffs are reset to zero.
\end{itemize}

While the networks we use are obtained from experimental measurements and,
as such, are given, there are different options for the learning rule of
the agents we place on the network. We chose to stick to the (unconditional)
imitation rule described above for a 
a number of reasons. From the
methodological viewpoint, imitation allows a direct comparison to
other studies, such as \cite{Nowak2,Nowak3} while, on the other hand, its
deterministic character makes its numerical study much more amenable.
Importantly, for global interactions learning by imitation ends up in
global defection, and hence cooperation in
a local model can not be due solely to this learning rule.
From the
theoretical viewpoint, it is clear that other rules, such as
best-reply, will lead straightforwardly to a fully defecting
population even with local interactions. It can be argued that
imitation is too simple a rule but, as discussed in \cite{Eshel},
there are several reasons why agents may fail to recognize they are
in a dilemma situation, which would lead them to defection; another
reason for the use of imitation is as a mode of economizing behavior
\cite{Pingle}.
From
the experimental viewpoint, there are several reports that indicate
that imitation is commonly used by humans
\cite{Kosfeld,Selten,Apesteguia}. Finally, imitation can be justified
in psychological terms by looking at how confirmation and
disconfirmation of beliefs are carried out
\cite{Strang} and has been also proposed as a relevant force
to drive the evolution towards economic equilibrium \cite{walras}.
Specific aspects where
the use of other learning mechanisms can change our results will
be discussed below (see Conclusions).

Finally, we note that the update rule for strategies is synchronous, i.e., all agents
update their strategy at the same time, proceeding to a new round of the game
subsequently. Changing to a non-synchronous update is known to have non-trivial
consequences \cite{Huberman,prl}. However, non-synchronicity
is difficult to deal with
in general, as the
particular way to introduce it comes dictated by the application of interest 
and different procedures lead to different results; that is why it
has been considered only rarely  in the framework of
evolutionary game theory, and only in very simple and arguably arbitrary ways.

\subsection{Results}
Let us begin by examining the results of simulations of the PD on
real social networks as a function of the temptation parameter $b$.
In Fig.\ \ref{figure1} we plot the final density of cooperators on the two cases
addressed here, the email network and the PGP network. The first remarkable feature of these
plots is the high level of cooperation attained even for large
values of $b$ on both networks, as compared to the results on
regular lattices \cite{Nowak2,Nowak3,Hauert} with the same
imitation dynamics. The cooperation levels are not as high as those
reported by Santos {\em et al.} \cite{Santos,santosprl,Santos3} 
on scale free networks,
although in their simulations the dynamics is stochastic, and
therefore a direct comparison can not be made. In this regard we also 
want to stress that
the two networks we are analyzing can not be considered scale-free: The email network has a clear exponential distribution of degrees, and the PGP network presents two regions with a clear crossover from a power law behavior with exponents $-2.63$ (for degree $k < 40$) and $-4$ (for degree $k > 40$) indicating strongly a bounded degree distribution.

Nevertheless, the crucial result arising from Fig.\ \ref{figure1} is 
that the
dependence of the level of cooperation on the temptation parameter $b$
is very different for both networks. As we may see from the plots,
the cooperation level on the email network is a decreasing function of
$b$, going from values very close to unanymous cooperation for
$b\gtrsim 1$ to about a 15\% for $b$ close to 2. On the contrary, the
PGP network presents an almost constant cooperation level, with a
variation of a 10\% at most in all the range of $b$ values, except
for $b=2$. These results inmediately lead to the conclusion that there
is no typical behavior of the cooperation level on true social
networks, at least in the framework of the PD with imitation dynamics
or learning.

The above conclusion is further reinforced by noting that the cooperation
level in each network changes in a very different manner when their
original structure is distorted. To this end,
we have compared the results on the two networks with their randomized version preserving the degree of each node, carried out through a rewiring process \cite{roberts}. The process,
that consists of repeatedly choosing at random two nodes and exchanging one neighbor of each node (also selected randomly), destroys correlations between nodes (and in particular the community structure we will discuss below). 
Figure \ref{figure1} shows clearly
that playing the game on the real networks
and on their randomized versions gives rise to opposite
behaviors: On the email network cooperation reaches
extremal values, higher than the random case when $b$
is close to 1, and lower when $b$ is close to its maximum limit of 2. On
the contrary, on the PGP network 
cooperation is higher on the random version for low
values of the temptation
$b$, and worse for higher values. Remarkably, the cooperation
level in the random versions of the two networks is very similar, and close to those reported in
\cite{Santos3} for the configuration (random) model, although
it must be kept in mind that the dynamics is different in the latter case; interestingly,
this does not seem to induce large differences in behavior in this respect.

\section{Discussion and hypothesis}

Our two examples, email and PGP, do not seem to fit in any of the categories previously reported in the literature for the behavior of the PD, which implies that
 the macroscopic (global, statistical) similarities between both topologies (see Table \ref{tab:nwk}) are not determinant for the opposite behaviors observed.
Furthermore, the fact that randomization, while preserving the degree distribution, drives the behavior of the two networks to the same general pattern, indicates
that neither the whole network nor individual agents
provide the clue to understanding our observations.  Therefore, in
order to gain insight on this problem,
we must consider an intermediate,
mesoscopic organizational level as the possible source of the explanation for the dramatic
differences observed in the original systems. This in turn requires
a deeper analysis of the structure of both networks, which is what 
we subsequently do. 

\subsection{Communities}
As the key concept to understand networks at a mesoscopic level, we propose to
focus on their community structure. Community structure is a common feature of
many networks: Communities can be qualitatively defined
as subgraphs with dense connections within
themselves and sparser ones between them, and very generally have functional
implications \cite{Newman}. More quantitatively, communities are introduced
through optimizing the quality function known as modularity: $Q=\sum_r(e_{rr}-a_r^2)$, where $e_{rr}$ are the fraction of links that connect two nodes
inside the community $r$, $a_r$ the fraction of links that have
one or both vertices inside the community $r$, and the
sum extends to all communities $r$ in a given network. 
The modularity of a given partition is then
the probability of having edges falling within groups in the network minus the expected probability in an equivalent (null case) network with the same number of nodes, and edges placed at random preserving the nodes' degree. 
The community
distribution of the network is then the partition that maximizes the modularity.
Among the wide
variety of algorithms available to carry out this maximization
process \cite{Danon}, we have chosen a divisive algorithm based
on Extremal Optimization (EO) heuristics \cite{Boettcher}. A
detailed description of the method is beyond the scope of the paper, but full details can be found elsewhere \cite{Duch}.

Once we have determined the number and size of the network communities,
we focus on the study of two structural mesoscopic characteristics:
The connectivity between communities and their internal organization.

\subsection{Inter-community structure}
To summarize the results obtained from a community analysis of both
social networks and to facilitate their comparison,
the outcome of our analysis is jointly presented in
Fig.\ \ref{figure2}{\em A} and Fig.\ \ref{figure2}{\em B} for the email and PGP respectively. Each node corresponds to a community, and a link between two
nodes denotes cross-relations. In addition, the size of nodes and
links gives information about community size and number of
cross-links, respectively. It is evident from the plot that communities in the email
network are densely interconnected, and sparsely interconnected in the PGP
network. The calculation of the
weighted degree distribution (the distribution of the sums of weights of links
for each node) $P(w)$ confirms this evidence: the email community network has a $P(w)\sim \exp^{-\alpha w^{2}}$ while the PGP community network 
presents a $P(w)\sim \exp^{-\beta w}$.

\subsection{Intra-community structure}

The internal structure of communities in both networks also presents
important
differences. 
In Fig.\ \ref{figure2}{\em C} and Fig.\ \ref{figure2}{\em D} we plot the aspect of representative communities of the email and PGP networks, respectively. From the plot, the differences in the internal structure are clear: the email communities present a very homogeneous structure when compared with the heterogeneity of 
the PGP communities. For a more quantitative assesment of this difference,
we have calculated the relative difference between the average and the
maximum value of the internal degree in each community, $\Delta H$. This measure allows to grasp the heterogeneity of the internal community structure. While $\Delta H \sim 5$ in the email network, the values of $\Delta H$ in the PGP network range from $5$ up to $35$, confirming our observations. In the following, 
we will call \emph{local hubs} the nodes in the PGP networks responsible for the very high $\Delta H\approx 30$.

\subsection{Discussion}
Previous works
have stressed the role of hubs at a macroscopical level in PD dynamics
on adaptive networks
\cite{Eguiluz,Eguiluz2,otropacheco,yamir}. Although our networks are static,
it is expected
that the presence of these local
hubs in PGP communities (as well as their absence in the email ones)
influences strongly the evolution of the PD on these networks. To be specific,
local hubs play a double stabilizing role: First,
as most nodes in the community are directly linked to their
local hub, the whole community tends to imitate the strategy of the
hub; second, when a less connected member of the community changes her strategy
following an external node, the influence of the local hub
makes it harder for this strategy to spread to the whole community. 

On the
contrary,
homogeneous internal degree distributions, as in the case of the email network,
lead to a behavior
that is not governed by hubs: All nodes are more
or less equivalent, and indeed simulations show that
their strategies evolve in a synchronized manner,
at least to some degree. Therefore, the behavior of the email network will
be more dependent on how the communities are connected among themselves. We thus
are in a position to formulate our hypothesis: the behavior observed in a
network with communities depends strongly on the intra-community heterogeneity
(IH) and on the inter-community connectivity (IC). In this scenario, the
robustness of cooperation observed in the PGP network is due to its low IC
and high IH, whereas the fact that cooperation only arises for low $b$ in
the email network arises from its high IC and low IH.

\section{Test of our hypothesis: model networks}

\subsection {Synthetic network model}

As we have seen, 
the analysis of the email and PGP networks raised two characteristic
patterns of the mesoscale: (i) IH, or existence or not of local hubs
in the  network, and (ii) IC, the degree of connections between
communities. To test this hypothesis, we propose to use synthetic
networks as a benchmark in which to tune the above mechanisms as
follows:
\begin{itemize}
    \item First we divide a number of nodes $N$, into $m$ communities of equivalent size.
    \item Second, we prescribe the IH. We have used as standard mechanisms for the construction of \emph{ad hoc} homogeneous and heterogeneous communities
the Erdos-Renyi model\cite{ER} and the heterogeneous (scale-free)
random graph resulting from the Barabasi-Albert model\cite{BA}, respectively. In
the first case the probability of connection between two nodes is
constant ($p_{intra}$); in the second case, the network grows by adding nodes 
with $k_0$ links to an
initial connected core, and the probability of connection of a node
$i$ to another existing node $j$ is proportional to the current
degree of node $j$.
    \item Third, we prescribe the IC. We construct a unique connected component by linking the communities previously generated. To interconnect the resulting communities we prescribe a new constant probability $p_{inter}$ to form links between two randomly selected nodes from the pool of communities, whenever these nodes below to different communities. The density of cross-connections is controlled by the probability $p_{inter}$. Note that $p_{inter}$ must be sufficiently large to ensure the existence of a unique connected component, but not so high as to mask the actual communities (i.e. an accurate detection algorithm should still separate the prescribed communities).
    \item Finally, we check by using a community detection algorithm (extremal optimization \cite{Duch}) that the communities obtained at the best partition of modularity are the prescribed ones.
\end{itemize}

We have built up four statistically significant synthetic test
networks with the same number of nodes ($N=10000$) and the same number
of communities ($m=75$), corresponding to four extremal
configurations corresponding to the combination of low and high
values of the IH and IC. Our expectation is that the configuration
corresponding to low IH and high IC will be representative of the
class of mesoscopic traits observed in the email network;
conversely, high IH and low IC should be representative of the class
of mesoscopic traits observed in the PGP network. The other two
cases, low IH and low IC, and high IH and high IC should constitute
intermediate configurations between the former ones. The statistical
properties of the so obtained networks are listed in Table
\ref{tab:new}.

At this point, we want to emphasize that the statistical
properties of the email and PGP networks (Table \ref{tab:nwk}) and
their synthetic counterparts (cases $A$ and $D$ in Table
\ref{tab:new}) present strong dissimilarities. First, we notice that
the degree distributions of the synthetic networks are different
from those observed in real networks. In addition, the clustering
coefficient of the synthetic networks is almost an order of
magnitude smaller than in the real networks. Finally, the
assortativity coefficient of the synthetic class with low IC 
is negative, while the other two present positive values
of assortativity. These different statistical properties of our
synthetic and empirical networks are specially interesting for the
validation process: Since the list of similarities between the two
sets of networks has been reduced to the desired inter and
intracommunity structural properties, any agreement we may find on the behavior
of cooperation dynamics can be safely attributed to these mesoscopic
features.


\subsection{Results}
Figure \ref{figure3} shows the evolution of cooperation as a function of the temptation parameter $b$ for our four synthetic networks. 
We discuss first the behavior of networks corresponding to A and D configurations (the \emph{synthetic classes} of the empirical email and PGP networks,
respectively). Although, in general, the values of final density of
cooperators are smaller than those reached in the empirical
cases (see Figure \ref{figure1}), synthetic networks reproduce the qualitative behaviors of
the two real social networks in terms of sensitivity to changes of
the temptation value $b$. Actually, the evolution of cooperation on the synthetic networks
presents the observed tendencies even more emphasized than the empirical
ones. On the one hand, all values of density of cooperators in case D
(\emph{PGP-class}) are close to the density established as initial
fraction of cooperators ($\rho =0.5$), revealing extraordinarily high levels of stability
of the strategies played by agents. On the other hand, the decrease
on the cooperation level shown in case A (\emph{email-class}) is
somewhat larger than that of the empirical email
network for the same range of temptation values.

Additional plots in Fig.\ref{figure3} help us to understand,
separately, how each mesoscopic characteristic acts over
cooperation. Comparing the behavior of configurations A and D with
the other two classes, B and C, we observe that when both mesoscopic
characteristics are high (configuration B), the system presents
remarkable rates of cooperation. On the other hand, for low values of
both mesoscopic characteristics (configuration C) the sensitivity to
the temptation parameter is increased, the maximum level of cooperation 
is smaller, and the decay on cooperation
is sharper. Consequently, we observe that IH seems to be more
determinant than IC as a stabilizing factor against changes on the
temptation to defect. Conversely, IC is more relevant to the maximum
level of cooperation reached than IH. It is then clear that the
behaviors observed on the email and PGP empirical networks (and on
their synthetic counterparts) is the result of the interplay of both
mesoscopic structural properties, since we cannot reach case D from
case A by tuning only one of them.

\section{Conclusions}

In this work we have addressed the issue of the emergence of cooperation
on true social networks in the framework
of the evolutionary PD with imitation.
Our results on two different networks show clearly
that the specific details of the network considered are very relevant to
determine the level of cooperation reached. Our analysis
of the
community structure of both networks lead us to the hypothesis that two mesoscopic structural properties
(the connectivity between communities, IC, and their internal structure, IH)
influence the evolution of cooperation in social networks by
raising or lowering the level of cooperation and the
stability of the behavior of the communities
against changes on the temptation to defect. In order to verify this claim,
we have designed synthetic model networks where these two features can
be tuned as desired. Simulations on four such synthetic networks confirmed
that, though their
structural features have
little in common with the empirical ones, except for the mesoscopic
characteristics under study, the behavior of
cooperation is very similar. Our models also show that both mesoscopic structural
characteristics, IH and IC, influence the robustness of cooperation against
changes on the temptation to defect, in excellent agreement with the observation made in the real social networks analyzed.

We stress that, as stated in the introduction, our results combine two ingredients:
locality (given by the network) and learning by imitation. In this paper we
focus on the network structure and find uncontestable evidence of the relevance
of IH and IC on the dynamics given by our update rule, unconditional
imitation. This is enough to claim that network structure has to be taken into
account in general, as aggregate characteristics may not give clues to understanding their behavior. However, we realize that the question then
arises as to the influence of these network
features on other dynamics. A thorough study of this issue is beyond
the present work, because evolutionary game theory on graphs depends very
strongly on the specific rule considered, and there are very many
different choices \cite{Hauert,Szabo,nosotros}. 
In the case of the networks studied
here, it is important to have in mind that 
unconditional imitation leads to {\em lower} levels of cooperation
\cite{nosotros}
than the stochastic rule used in \cite{Santos} (proportional update). On the 
other hand,
hubs have been shown recently to play a role similar to the one
discussed here under such a proportional update dynamics \cite{yamir}. 
We then envisage that, at least for that specific rule our results will apply
more or less straightforwardly, with even higher cooperation levels. 
On the other hand, best-response-type rules
lead, generally speaking, to the same outcome as well mixed populations
\cite{nosotros}, and it is clear that in that case the network structure
might control the time to reach asymptotics, but not the final state itself.
In any event, it is clear that this issue deserves further and thorough study. 

Dwelling further on the
evolutionary perspective, the work by Egu\'\i luz {\em et al.}\
\cite{Eguiluz} indicates that if the network is allowed to co-evolve with
the strategies, a network with hubs develops. Interestingly, in this network
with hubs, the cooperation level shows similar dependence on the temptation
parameter, much as we have found here for the PGP network. Along similar
lines, recent work by Santos {\em et al.} \cite{otropacheco,nuevopacheco} suggests
a connection between the emergence of cooperation and the evolutionary
appearance of degree heterogeneity. In this context, our study, which
we stress is carried out on static networks, suggests that the cooperation
levels we observe in the PD may be related to the different origin of
the two networks: While the PGP network is spontaneously formed and with
a clearly cooperative goal in mind (namely, finding help to ensure
communication privacy), the email network arises from an underlying
external structure, whose main purpose is not so clearly cooperative as
it involves many other aspects and tasks. Our results would then
support the existence of community structures organized around hubs
with resilient cooperative behavior.

\begin{figure}
\begin{center}
 \epsfig{file=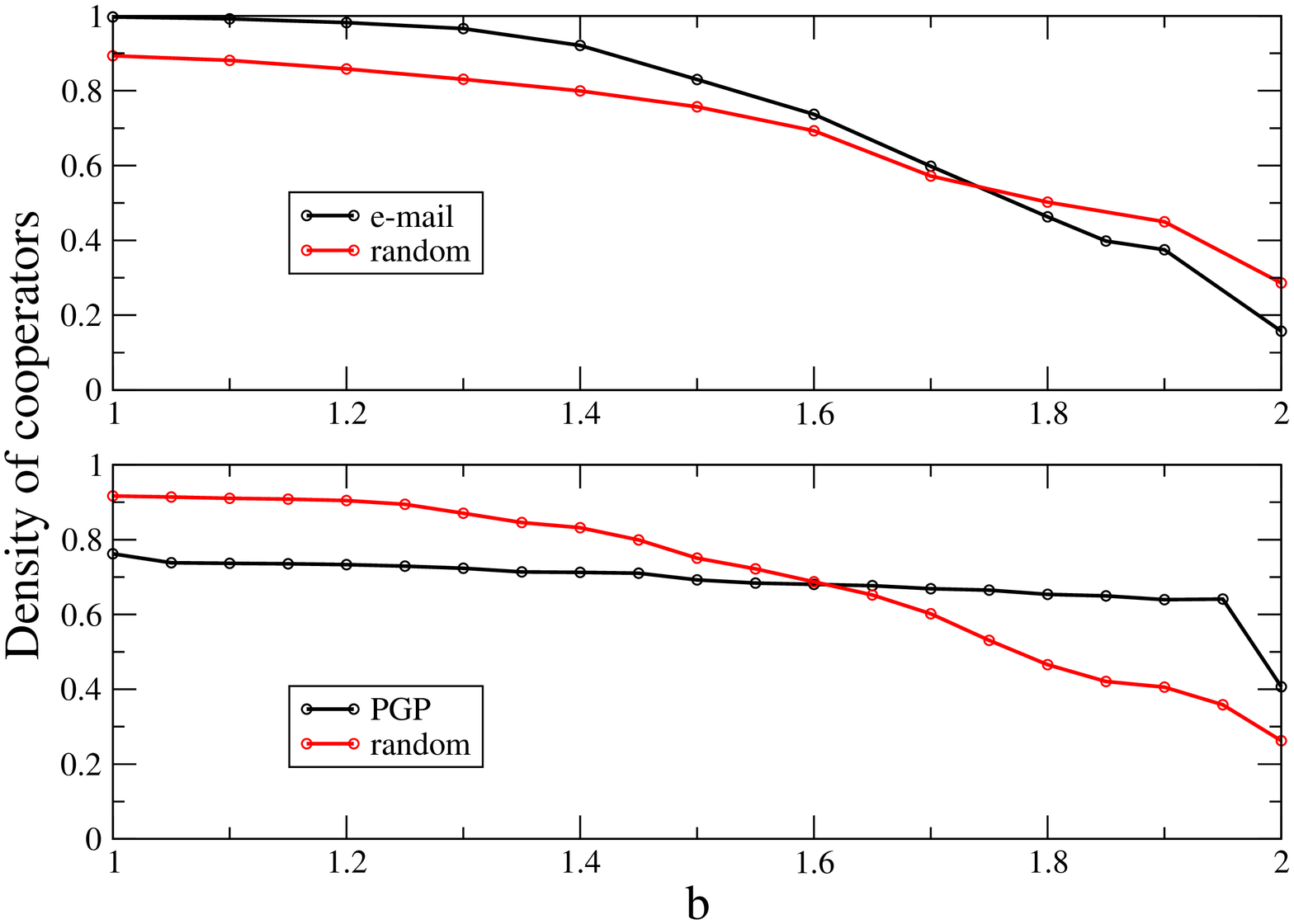, width=12cm, angle=270}
 \end{center}
 \caption[Evolution of cooperation in real social networks.]%
 {Evolution of cooperation in real social networks. Black lines:
 Density of cooperators as a function of $b$, obtained by
numerical simulations on the email (top) and PGP (bottom) networks.
Red lines: Density of cooperators on random networks generated from the
original ones by a rewiring procedure that preserves the degree
distribution(see text). The
equilibrium densities of cooperators have been obtained by averaging
500 generations, after a transient time of 750 generation steps.
Each point corresponds to an average over 1000 independent
simulations with 50\% cooperators and defectors as the initial
condition.}
\label{figure1}
\end{figure}

\begin{figure}[!tpb]
  \begin{center}
  \begin{tabular}[t]{cc}
    \multicolumn{1}{l}{\bf A}
    &
    \multicolumn{1}{l}{\bf B}
    \\ \\
    \mbox{\includegraphics*[width=.5\textwidth]{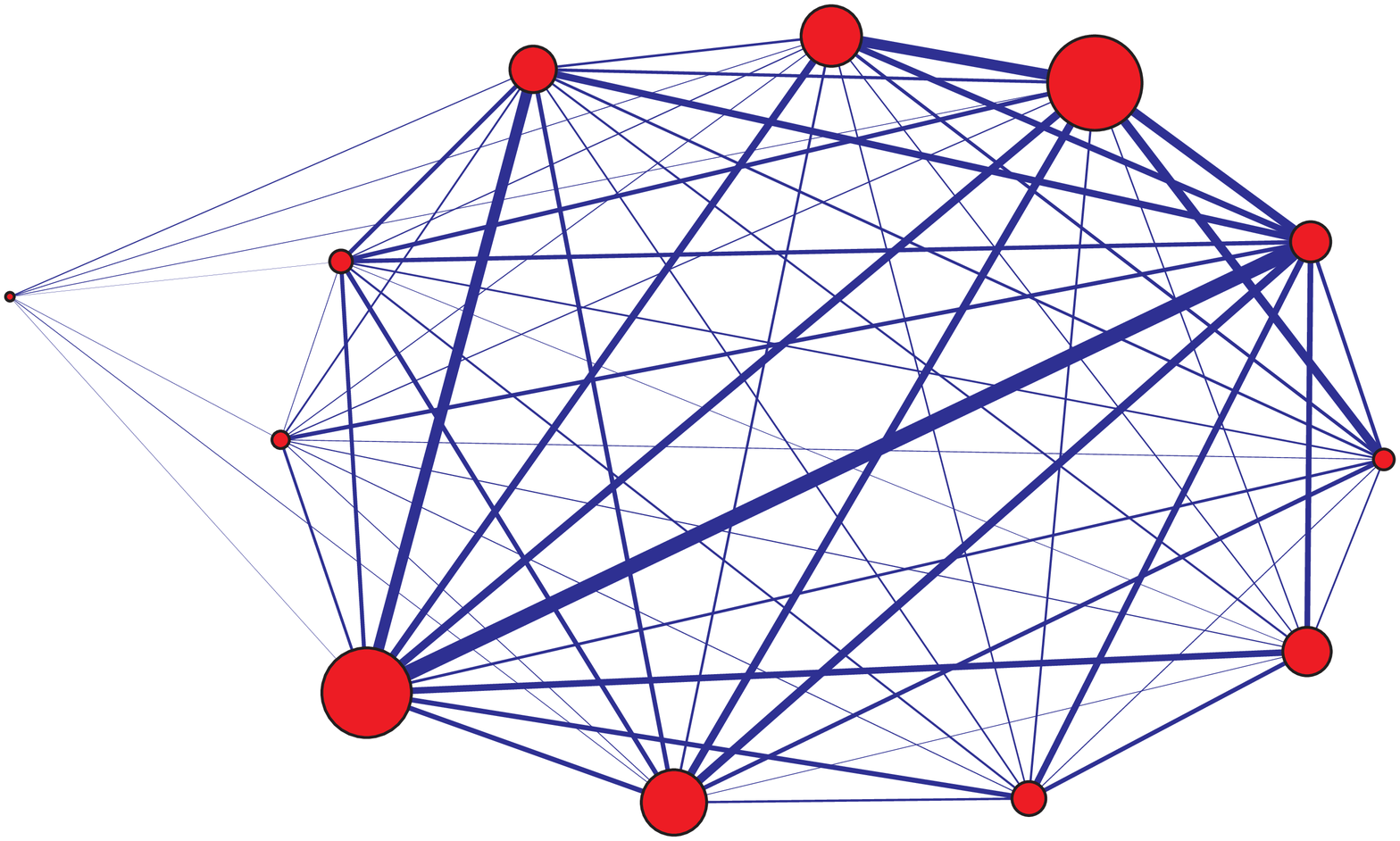}}
    &
    \mbox{\includegraphics*[width=.5\textwidth]{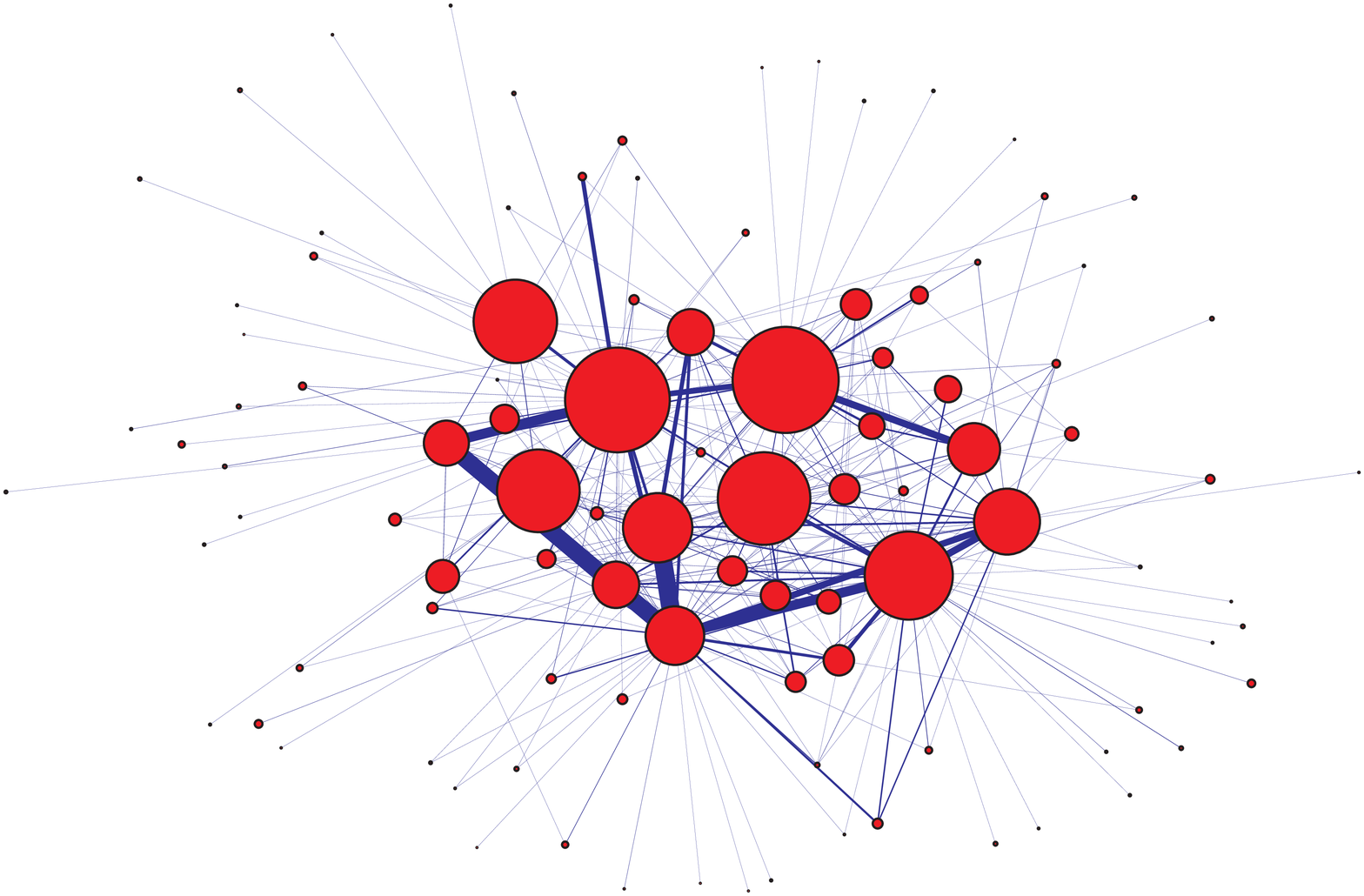}}
    \\ \\
     \multicolumn{1}{l}{\bf C}
    &
    \multicolumn{1}{l}{\bf D}
    \\ \\

    \mbox{\includegraphics*[width=.5\textwidth]{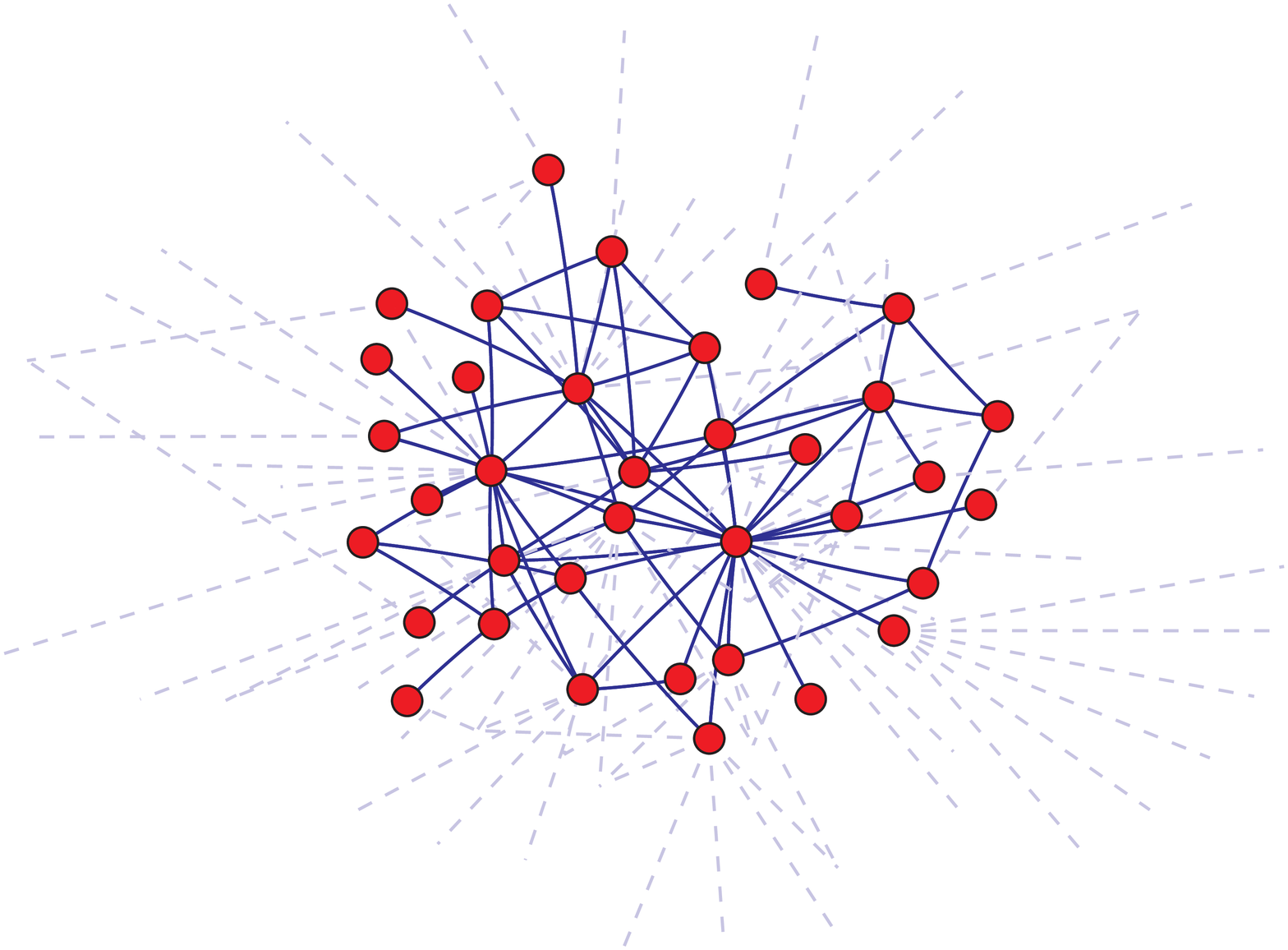}}
    &
    \mbox{\includegraphics*[width=.5\textwidth]{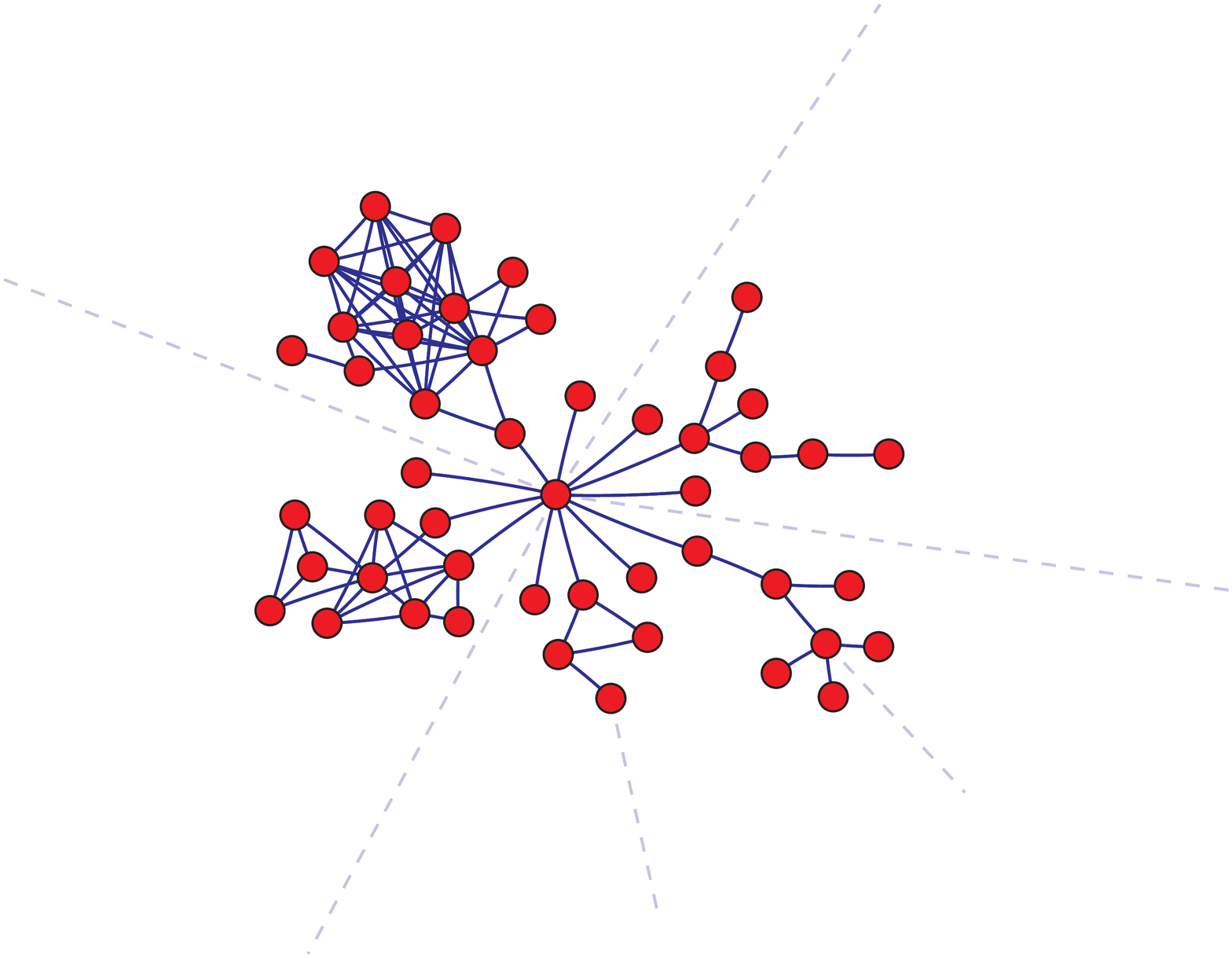}}
  \end{tabular}
  \end{center}
\caption{Community structures of the email and PGP
networks. Top: Community structures of the email (A) and PGP
 (B) networks.
Nodes correspond to communities (where size is proportional to their
number of members) and links represent cross-connections (where
width corresponds to the number of inter-connetions). Bottom:
Typical examples of the communities detected in the email (C) and PGP
(D) networks. Solid links join nodes of the community, dashed links join this community with others.}
\label{figure2}
\end{figure}

\begin{figure}
\begin{center}
 \epsfig{file=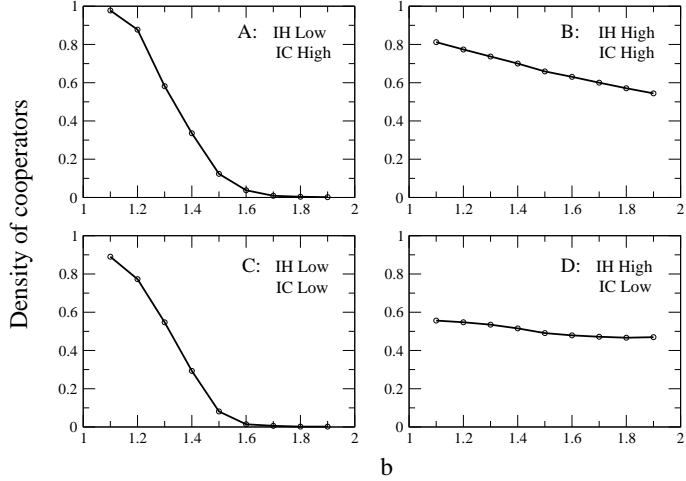, width=8cm, angle=270}
\end{center}
 \caption[Evolution of cooperation on synthetic networks.]{Evolution of cooperation in four synthetic networks. Cases $A$ and $D$ correspond, respectively, to the synthetic classes of networks akin to the email and PGP real networks.
 In case $A$ communities have been built as Erdos-Renyi random graphs ($p_{intra}=1.5\e{-1}$), and the probability of interconnection between communities ($p_{inter}$) is $5\e{-2}$. Communities in case $D$ are constructed as independent scale-free networks (Barabasi-Albert with $k_0=3$), and after they have been sparsely interconnected  with ($p_{inter}=1.5\e{-5}$). Case $B$ has been obtained from $D$ by increasing the probability $p_{inter}$ to $3.5\e{-4}$, and case $C$ corresponds to $A$ reducing this probability to $7.5\e{-4}$. Simulations have been performed as indicated in Fig.\ 1.}
 \label{figure3}
\end{figure}

The above comment suggests, in addition, that our results may be of
interest for the design of hierarchies and organizations with
tailored cooperation behavior. We have seen that the email network
reaches, for moderate values of the temptation parameter,
cooperation levels very close to the optimum. Therefore, networks
with this structure should be used in order to achieve very high
performance levels in terms of cooperation. On the other hand, while
the email network is quite susceptible to an increase of the
temptation parameter, and hence exhibits a degrading of the
cooperation for large temptations, the PGP network, with its weakly
connected communities with hubs, is much more robust in this
respect, and ensures cooperation for almost any temptation.
Organizations with a PGP-like structure would exhibit a very robust
cooperation, although there would always be defectors. Further
research at the mesoscopic scale, looking at different combinations
of IH and IC structures, could lead to designs that would be both
optimal and robust (such as, e.g., the structure in Fig.\
\ref{figure1}B). Interestingly, this conclusion may carry over to 
different dynamical contexts (other than evolutionary game theory):
For instance, recent results on synchronization dynamics in a 
system of coupled oscillators show a strong influence of the 
community structure as well \cite{alex}, and hence communities have to 
be taken into account much in the same way we are describing here. 

Finally, we want to emphasize our main conclusion, namely
that cooperation in real social
networks is a complex issue depending on the combination of the
effects of several structural features. This result has far-reaching implications:
Thus,
several previous researches have considered how cooperation emerges
in the PD on different model networks, including gaussian, scale free and
small world ones as paradigms of social networks. There are two
main differences between our work and those previous ones: first,
the cooperation level is in general higher that in the model networks, and
second, results are very different for similar global parameters
of the network due to the influence of the community structure, often
undetected by global measurements. It is then clear
that any approximation to the evolution of cooperation in social
networks based on the generalization of only one of these structural
features is far too simplistic and may be misleading. We envisage that
similar conclusions may apply to other models of cooperation or
coordination based on other games, as arguments based on the inter-
and intra-structure of the communities may well carry over to them.
In any event, we believe that subsequent studies on
these issues should then be carried out on a case by case basis, and
should involve a careful analysis at a mesoscopic (community) level,
trying to find out whether behaviors can be predicted or classified in
classes attending to this structure.

\begin{acknowledgments}
We thank Carlos P. Roca and Jos\'e A. Cuesta for help with the computer
simulations, and Esteban Moro and Yamir Moreno for discussions. This
work is supported by Ministerio de Educaci\'on y Ciencia (Spain) under grants
FIS2006-13321, MOSAICO and NAN2004-9087-C03-03 and by Comunidad de Madrid
(Spain) under grants
UC3M-FI-05-007 and SIMUMAT-CM. S.L. is supported by URV through a FPU grant.
\end{acknowledgments}



\end{article}




\mbox{}

\newpage

\begin{table}
  \caption[Statistical properties of e-mail and PGP networks.]{Statistical properties of e-mail and PGP networks. N is the number of nodes of the giant component of the network considering only those links that are bidirectional (indicating mutual acquaintance between nodes). $P(k)$ is the degree distribution (best fit to the data). $<C>$ is the clustering coefficient, and $r$ stands for the assortativity coefficient \cite{newmanassor}.\label{tab:nwk}}
  \begin{center}
\begin{tabular}{cccccc}
  \hline
  \textbf{ network } & \textbf{ ref. } & \textbf{\hspace{15pt}N\hspace{15pt}} & \textbf{ P(k) } & \textbf{$ <C> $} & \textbf{\hspace{15pt}r\hspace{15pt}} \\
  \hline
  \hline
  email & \cite{Guimera} & 1133 & $\sim \exp^{-k/9.2}$ & 0.25 & 0.078 \\
  \hline
  PGP & \cite{Boguña} &  10680  & $ \sim \left\{ \begin{array}{ll} k^{-2.63} &\textrm{if $k<40$}\\ k^{-4.0} &\textrm{if $k>40$}\\ \end{array}
  \right. $ & 0.26 &  0.238 \\
  \hline
\end{tabular}
\end{center}
\end{table}

\begin{table}
\caption[Statistical properties of synthetic networks.]{Statistical properties of synthetic networks with 10000 nodes and 75 communities. $P(k)$ is the degree distribution (best fit to the data), 
$<C>$ is the clustering coefficient, and $r$ stands for the assortativity coefficient \cite{newmanassor}.\label{tab:new}}\begin{center}
\begin{tabular}{lcrr}
\multicolumn1c{$Class$}&\multicolumn1c{ P(k)}&
\multicolumn1c{$<C>$}&\multicolumn1c{{\bf r}}\cr
\hline
\hline
A (Low IH - High IC)&$\sim \exp^{-0.0018k^{2}}$&0.031&0.013\cr
\hline
B (Low IH - Low IC)&$\sim \left\{ \begin{array}{ll} k^{-2.47} &\textrm{if $k<30$}\\ \exp^{-0.047k} &\textrm{if $k>30$}\\ \end{array} \right.$& 0.040& -0.202\cr
\hline
C (High IH - High IC)&$\sim \exp^{-0.003k^{2}}$& 0.080&0.110\cr
\hline
D (High IH - Low IC)& $\sim \left\{ \begin{array}{ll} k^{-2.38} &\textrm{if $k<30$}\\ \exp^{-0.027k}  &\textrm{if $k>30$}\\ \end{array}
 \right.$& 0.090&-0.308\cr
\hline
\end{tabular}
\end{center}
\end{table}




\end{document}